\documentclass[pra,twocolumn,showpacs]{revtex4}
\usepackage{graphicx,amsmath}

\begin{document}
\date{\today}

\title{Dissipation effect in the double-well Bose-Einstein Condensate}

\author{Hanlei Zheng, Yajiang Hao and Qiang Gu\footnote[1]{Email: qgu@ustb.edu.cn}}

\address{Department of Physics, University of Science and Technology
  Beijing, Beijing 100083, China}

\begin{abstract}
Dynamics of the double-well Bose-Einstein condensate subject to
energy dissipation is studied by solving a reduced one-dimensional
time-dependent Gross-Pitaevskii equation numerically. We first
reproduce the phase space diagram of the system without dissipation
systematically, and then calculate evolutionary trajectories of
dissipated systems. It is clearly shown that the dissipation can
drive the system to evolve gradually from the $\pi$-mode quantum
macroscopic self-trapping state, a state with relatively higher
energy, to the lowest energy stationary state in which particles
distribute equally in the two wells. The average phase and phase
distribution in each well are discussed as well. We show that the
phase distribution varies slowly in each well but may exhibit abrupt
changes near the barrier. This sudden change occurs at the minimum
position in particle density profile. We also note that the average
phase in each well varies much faster with time than the phase
difference between two wells.

\end{abstract}


\maketitle

\section{Introduction}

The experimental realization of atomic Bose-Einstein condensation
has stimulated great interest in quantum nature of the condensate of
weakly interacting bosons. Since the condensate is in the
macroscopic quantum state, a variety of quantum phenomena show up
when two condensates are linked. If the two condensates are
separated by a barrier through which atoms can tunnel, a bosonic
Josephson junction (BJJ) is built and allows for the detection of
Josephson-like current-phase effects, analogous to two weakly
coupled superconductors. This phenomenon was initially predicted by
Javanainen \cite{Javanainen1986} and was further demonstrated by a
number of researchers using various techniques
\cite{Walls1996,Leggett1998,Smerzi1997,Smerzi1999,Walls1997,Collett1998,Vardi2001,Cederbaum2009,Kroha2009,Vardi2009,
Polls2010}. In addition, the dc and ac Josephson effects
\cite{Walls1996,Leggett1998,Smerzi1997,Smerzi1999} as well as the
Shapiro effect \cite{Smerzi1999} in BJJs have been discussed.

Furthermore, many theoretical works reveal that the BJJ can exhibit
new phenomena that are not accessible with superconductor Josephson
junctions (SJJs). The mean-field theory based on two-mode
approximation predicted the $\pi$-phase oscillations and macroscopic
quantum self-trapping (MQST) \cite{Smerzi1997,Smerzi1999}. The MQST
is a fascinating nonlinear phenomenon which arises from interactions
between atoms. Quantum corrections to the mean-field results were
also intensively discussed
\cite{Walls1997,Collett1998,Vardi2001,Cederbaum2009,Kroha2009,Vardi2009,
Polls2010}, providing a number of fascinating observations, such as
collapse and revival of quantum oscillations \cite{Walls1997},
occupancy of phase-squeezing state \cite{Collett1998}, disappearance of
coherence \cite{Vardi2001,Cederbaum2009}, and destruction of the
self-trapped state \cite{Kroha2009}.

In above studies the double-well condensate is treated as an
isolated system, but actually it is coupled to certain thermal cloud
and subject to environment distortions
\cite{Walls1998,Sols2003,Wimberger2008,Diehl2008}. Ruostekoski and
Walls studied the energy dissipation in BJJs via stochastic
simulations in a two-mode approximation and showed that the MQST
decays away \cite{Walls1998}. Wimberger {\it et al.} discussed
dynamics of the double-well condensate subject to phase noise and
particle loss and indicated that coherence might be increased to a
maximum for proper dissipation \cite{Wimberger2008}. They argued
that this phenomenon can be understood as a stochastic resonance of
the many-particle system.

On experimental side, the MIT team carried out the first double-well
experiment and gained interference patterns proving the coherence of
BEC \cite{Andrews1997}. In 2005, Albiez {\it et al.} reported the
direct observation of the quantum tunneling and macroscopic
self-trapping for a single BJJ \cite{Oberthaler2005}. They measured
the density distribution \textit{in situ} and deduced the relative
phase versus time from interference fringes. A summary of the
observed static, thermal and dynamical properties was given in Ref.~
\protect\cite{Oberthaler2007}. After that the dc Josephson effect
was also achieved \cite{Levy2007}. A classical bifurcation was
studied at the transition from the Rabi to Josephson oscillations
\cite{Oberthaler2010}.

In this paper, we concentrate on dynamical behaviors of the
double-well condensate subject to energy dissipation based on the
time-dependent Gross-Pitaevskii (TDGP) equation. The TDGP equation
provides a very good description of dynamical evolution of trapped
dilute Bose gases at temperature far below the critical temperature
\cite{Bergeman2006,Sanpera2011}. The two-mode approximation
for the BJJ is an extremely simplified version of the TDGP equation.
Although it can give a qualitatively description of the BJJ system
\cite{Walls1996,Smerzi1997,Walls1998,Wimberger2008}, the two-mode
model ignores the spatial distribution of the density and phase.
Whereas, these information could be obtained from the
Gross-Pitaevskii (GP) equation \cite{Bergeman2006}.

This paper is organized as follows. In Sect.~\ref{sect:model}, the
TDGP equation is introduced to describe the double-well condensate,
where a phenomenological term is added to account for the energy
dissipation. In Sect.~\ref{sect:diagram}, we reproduce numerically
the phase space diagram of the double-well condensate without
dissipation, and make a comparison with the two-mode model results.
Sect.~\ref{sect:dissipation} discusses the dissipation effect. A
summary is given in the last section.

\section{The model}
\label{sect:model}

\subsection{Time-dependent Gross-Pitaevskii (TDGP) equation}

The general formalism of three-dimensional (3D) time-dependent
Gross-Pitaevskii equation for dilute Bose gases trapped in the
double-well potential at zero temperature reads
\begin{eqnarray}\label{GP3D}
 \imath\hbar \frac{\partial\psi({\bf r};t)}{\partial t} = \left[
    -\frac{\hbar^2}{2m}\nabla^2+V({\bf r})+gN\lvert \psi({\bf r};t)
    \rvert^2 \right] \psi({\bf r};t),
\end{eqnarray}
where $\psi({\bf r};t)$ is the macroscopic wave function at position
$\bf r$ and time $t$, $m$ is the mass of an atom, $g=4\pi\hbar^2
a/m$ is the nonlinear constant with $a$ the $s$-wave scattering
length, and $N$ is the total atom number. The double-well potential
$V({\bf r})$ is given by $V({\bf r}) = \frac{m}{2}( \omega^{2}_x
r_x^2 + \omega^{2}_y r_y^2+\omega^{2}_z r_z^2) + V_{\rm b}
\mathrm{cos}^2(\pi r_z/q_0)$ \cite{Sanpera2011}, where the first
term describes a harmonic trap with trap frequencies
$\omega_{x,y,z}$, and the second denotes the barrier potential.
Hereinafter we set $\omega_x=\omega_y=\omega_{\bot}\gg \omega_z$.

Assuming that the dynamical evolution mostly takes place in the
direction containing the barrier, the 3D GP equation can be reduced
to a simpler one-dimensional (1D) form \cite{Bao2010}. Such
reduction is reasonable in the case of a strong trapping in the
other two spatial directions, say, in the $x$-$y$ plane. Then the 1D GP
equation in $z$ axis only with dimensionless quantities is given by \cite{Adhikari}
\begin{eqnarray}\label{GP1D}
 \imath\frac{\partial \bar\psi(z;\tau)}{\partial \tau} = \left[
    -\frac{\partial^2}{\partial z^2} +v_{1D}(z) +
    \frac{g_{1D}\beta}{\pi}\lvert \bar\psi(z;\tau)\rvert^2 \right]\bar\psi(z;\tau),
\end{eqnarray}
where $z=r_z/l_z$, $\tau=t\omega_z/2$ and
$\beta=\omega_{\bot}/\omega_z$. The dimensionless parameters are
scaled by the characteristic length $l_z=\sqrt{\hbar/(m\omega_z)}$
and energy $\hbar\omega_z$ correspondingly. The 1D potential is
reduced to $v_{1D}(z) = z^2+ 2v_{\rm b} \mathrm{cos}^2(\pi
z/q_0^{\prime})$, with $v_{\rm b}=V_{\rm b}/({\hbar\omega_z})$ and
$q_0^{\prime}=q_0/l_z$. $g_{1D}=gNm/(l_z \hbar^2)$ is the dimensionless
interaction parameter. This reduction greatly simplifies the
computation. The wave function $\bar\psi(z;\tau)$ satisfies the
normalization condition,
\begin{eqnarray}
\int^{\infty}_{-\infty}\mathrm{d}z\lvert\bar\psi(z;\tau)\rvert^2=1.
\end{eqnarray}

The reduced TDGP equation can be solved by a commonly accepted
numerical algorithm called Split-Step Crank-Nicolson scheme with
space and time both discretized \cite{Adhikari}. In numerical calculation, $t$ is
measured in $\omega_z/2$. If $\omega_z=2\pi\times 10Hz$, $\tau=1$
corresponds to $t=31.8ms$.

\subsection{The TDGP equation with dissipation}

Choi {\it et al.} has described a phenomenological damping theory
based on the TDGP equation \cite{Burnett1998}. A phenomenological
parameter $\gamma$ is introduced in the GP equation to account for
the dissipation rate. The reformed TDGP equation is written as
\begin{eqnarray}\label{GP1D-Dis}
 (\imath-\gamma)\frac{\partial \bar\psi(z;\tau)}{\partial \tau} &=& \left[
    -\frac{\partial^2}{\partial z^2} + v_{1D}(z) \right. \nonumber\\
    &+& \left. \frac{g_{1D}\beta}{\pi}\lvert \bar\psi(z;\tau)\rvert^2 \right]\bar\psi(z;\tau)~.
\end{eqnarray}
The phenomenological parameter $\gamma$ characterizes the strength
of dissipation and its value may be determined experimentally. Choi
{\it et al.} got that $\gamma =0.03$ by fitting theoretical results
with the MIT experiments dates. This approach is comparable to the
generalized finite-temperature GP equation derived by Zaremba,
Nikuni and Griffin \cite{Griffin1999},
\begin{eqnarray}
 \imath\hbar \frac{\partial\Psi({\bf r};t)}{\partial t} =
    [-\frac{\hbar^2}{2m}\nabla^2+V({\bf r})+gn+2g\tilde{n}-i\Gamma]\Psi({\bf r};t),
\end{eqnarray}
where $n({\bf r};t)=|\Psi({\bf r};t)|^2$ is the condensate density
and $\tilde{n}({\bf r};t)$ is the noncondensate density. Dissipation
arises from collisions between the condensate and noncondensate
atoms.

Ueda {\it et al.} have successfully employed this phenomenological
approach to study the detailed dynamics of a sudden rotated BEC in a
trap with $\gamma=0.03$ and figured out the vortex lattice formation
process \cite{Ueda2002}. This approach is also used to study the
Bose-Einstein droplet in free space \cite{Ueda2004}, vortex
formation in Bose-Einstein condensates in a rotating double-well
potential \cite{Wu2010} and the decay of dark solitons in
harmonically trapped, partially condensed Bose gases in the presence
of phase fluctuations or dynamical noise \cite{Cockburn2011}. These
works prove that the phenomenological damping description does
indeed provide an efficient numerical tool for the research of BECs
in an open system.

\subsection{The phase difference and phase distribution}

As discussed in the introduction, we consider two Bose-Einstein
condensates with repulsive interaction in a Quasi-1D symmetric
double-well potential, which could be experimentally realized by
splitting a single well into two parts: one cigar-shaped atomic
cloud is cut into two separated aligned cigars. It is necessary to
construct an approximate initial state of the double-well condensate
at first. We set $\phi_L$ for the wave packet mostly localized in
the left well and $\phi_R$ the right one. The original trial wave
function is forged as a superposition of $\phi_L$ and $\phi_R$,
\begin{eqnarray}
\bar\psi(z;\tau)=\psi_L(\tau)\phi_L(z)+\psi_R(\tau)\phi_R(z),
\end{eqnarray}
where $\psi_{L(R)}(\tau) = \sqrt{n_{L(R)}(\tau)}
\mathrm{e}^{i\theta_{L(R)}(\tau)}$. $n_{L(R)}(\tau)$ corresponds to
fraction of atoms on the left (right) side of the trap, and therefor
$n_{L}(\tau)+n_{R}(\tau)=1$. They are constructed as linear
combinations of the two lowest stationary modes of the 1D GP
equation, the symmetric $\phi_+$ and antisymmetric $\phi_-$.
$\phi_L(z)=\frac{\phi_+(z)+\phi_-(z)}{2}$ and $
\phi_R(z)=\frac{\phi_+(z)-\phi_-(z)}{2}$. $\phi_+(z)$ and
$\phi_-(z)$ can be obtained by the method of imaginary-time
propagation where the time variable $\tau$ in Eq. (\ref{GP1D}) is
replaced with an imaginary time.

The initial state with a given initial imbalance at $t=0$ is
constructed as follow
\begin{eqnarray}
\bar\psi(z;0) = \mathrm{e}^{i\Delta\theta(0)}\sqrt{n_L(0)}\psi_L(z) +
\sqrt{n_R(0)}\psi_R(z),
\end{eqnarray}
where $\Delta n(0)=n_L(0)-n_R(0)$ and $\Delta\theta(0)$ represent
the initial population imbalance and phase difference in the
double-well. The dynamical evolution is calculated by solving Eq.
(\ref{GP1D}) through real-time propagation method.

Once $\bar\psi(z;\tau)$ is obtained, the atom fraction in the left
and right well could be computed by
$n_L(\tau)=\int^0_{-\infty}\mathrm{d}z\lvert\bar\psi(z;\tau)\rvert^2$
and
$n_R(\tau)=\int_0^{\infty}\mathrm{d}z\lvert\bar\psi(z;\tau)\rvert^2$.
The population imbalance is $\Delta n(\tau)=n_L(\tau)-n_R(\tau)$.
The phase of condensates in the left well is defined as
\begin{eqnarray}\label{theta_L}
 \theta_L(\tau) = \mathrm{arctan} \frac{\int^0_{-\infty}\mathrm{d}z
    \mathrm{Im} [\bar\psi(z;\tau)]\rho(z;\tau)}{\int^0_{-\infty}\mathrm{d}z
    \mathrm{Re}[\bar\psi(z;\tau)]\rho(z;\tau)},
\end{eqnarray}
and in the right well
\begin{eqnarray}\label{theta_R}
 \theta_R(\tau) = \mathrm{arctan}\frac{\int_0^{\infty}\mathrm{d}z
    \mathrm{Im} [\bar\psi(z;\tau)]\rho(z;\tau)}{\int_0^{\infty}\mathrm{d}z
    \mathrm{Re}[\bar\psi(z;\tau)]\rho(z;\tau)},
\end{eqnarray}
where $\rho(z;\tau)= \bar\psi^*(z;\tau)\bar\psi(z;\tau)$. The phase
difference is $\Delta\theta(\tau)=\theta_L(\tau)-\theta_R(\tau)$.

Moreover, we could get the phase at a certain point at a certain
time which is different from that of the two-mode model.
\begin{eqnarray}\label{theta}
 \theta(z;\tau) = \mathrm{arctan} \frac{\mathrm{Im}\left[ \bar\psi(z;\tau)\right ]}
    {\mathrm{Re} \left[\bar\psi(z;\tau)\right]}.
\end{eqnarray}

\section{The phase space diagram}\label{sect:diagram}

\begin{figure}[htb]
\includegraphics[width=0.75\linewidth]{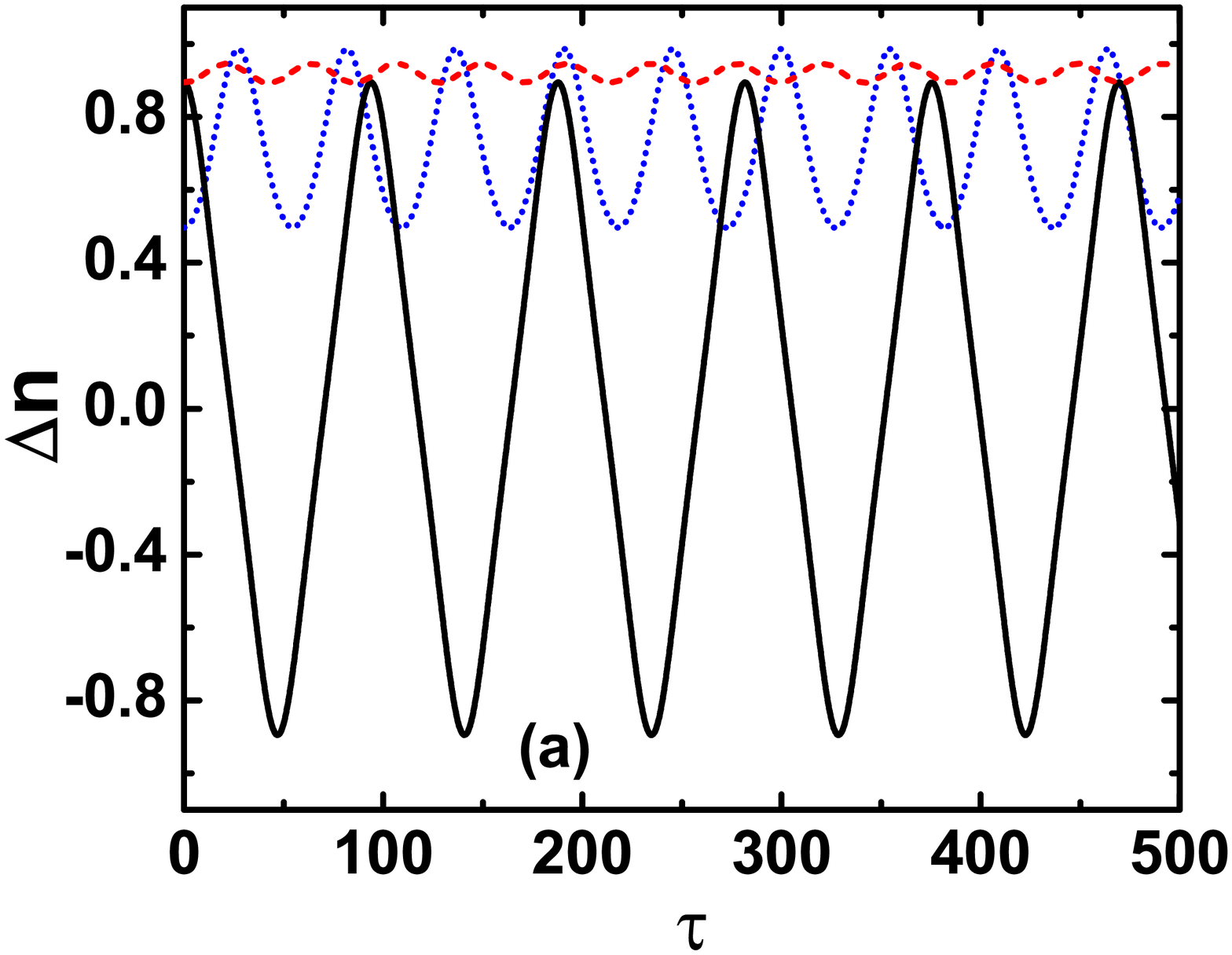}
\includegraphics[width=0.75\linewidth]{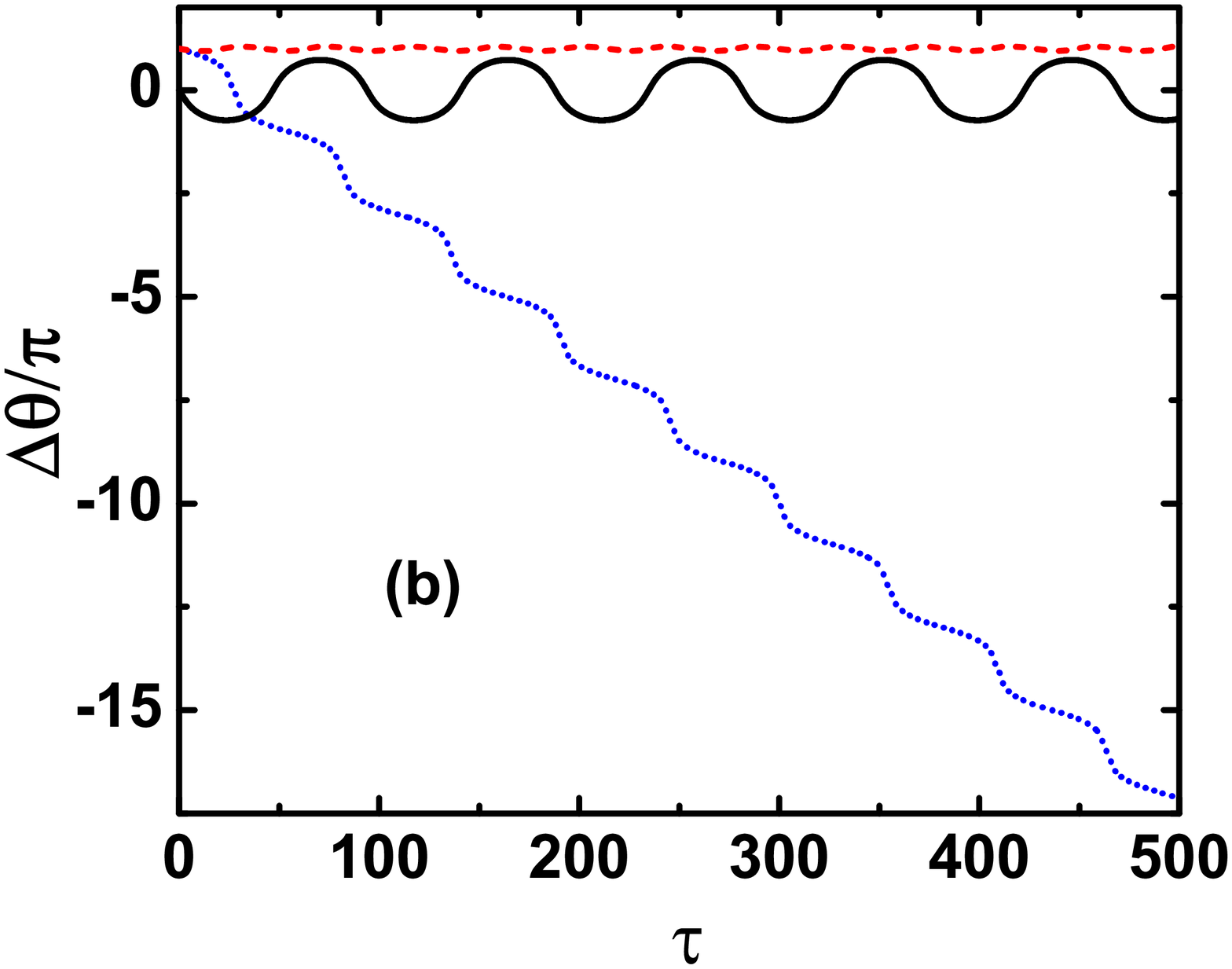}
\caption{(Color online) The $\Delta n$-$\tau$ plot (a) and $\Delta\theta$-$\tau$
plot (b). The solid lines, short-dashed lines, and short-dotted lines represent the evolutions from the initial state
($\Delta n=0.9$ $\Delta\theta=0$), ($\Delta n=0.9$ $\Delta\theta=\pi$) and ($\Delta n=0.5$
$\Delta\theta=\pi$), respectively.
} \label{fig:MQST}
\end{figure}

\begin{figure*}[htb]
\includegraphics[width=0.3\linewidth]{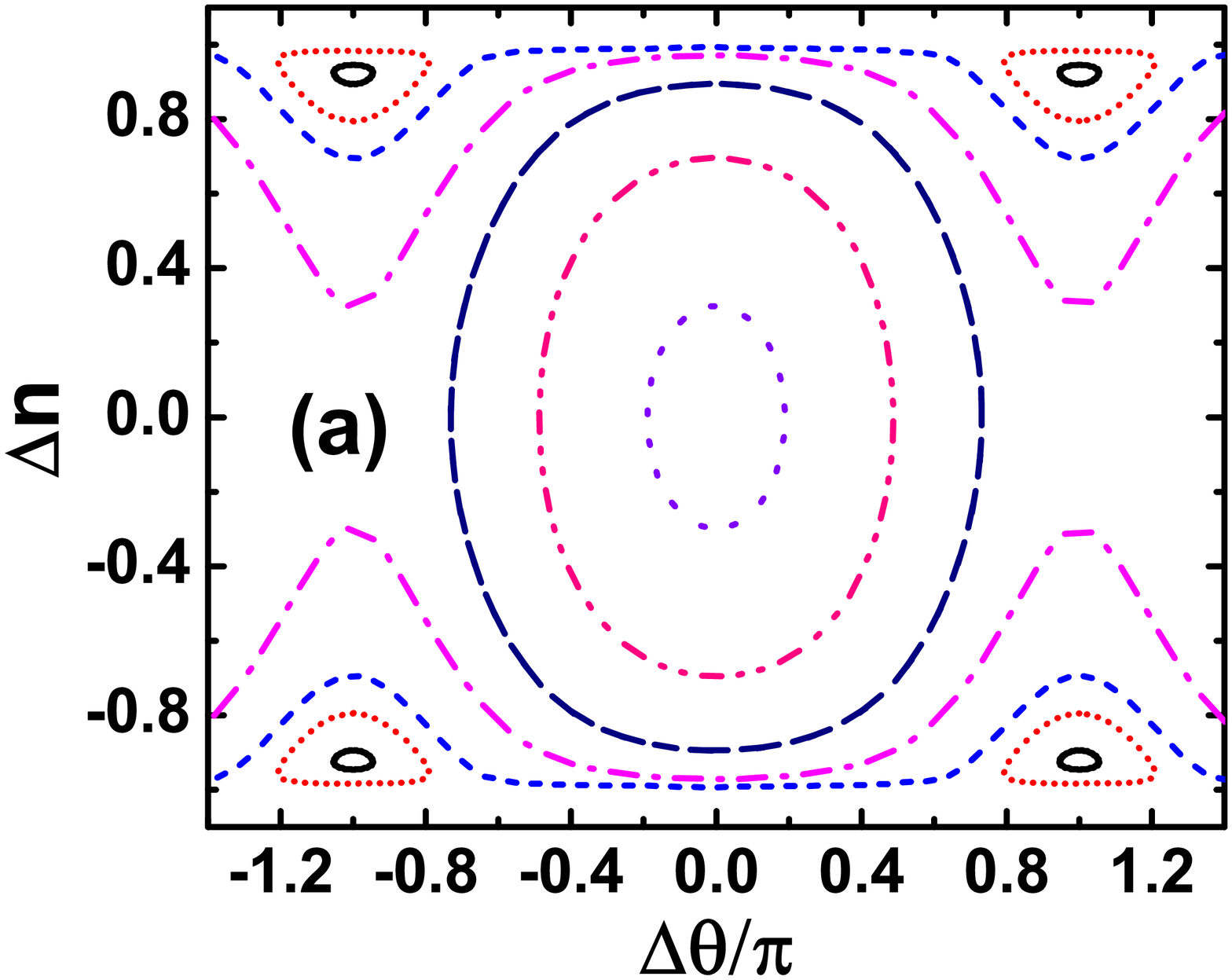}~~
\includegraphics[width=0.3\linewidth]{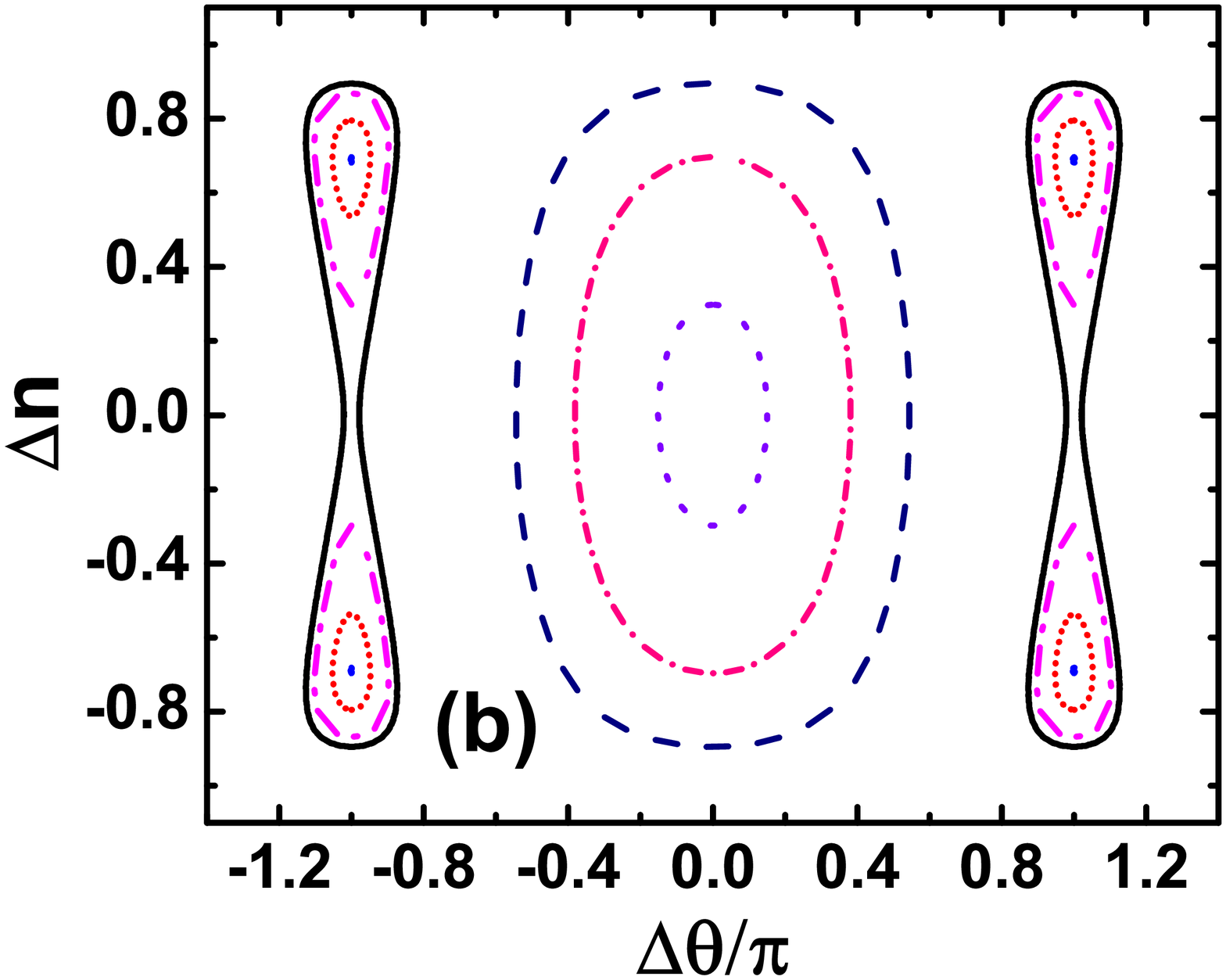}~~
\includegraphics[width=0.3\linewidth]{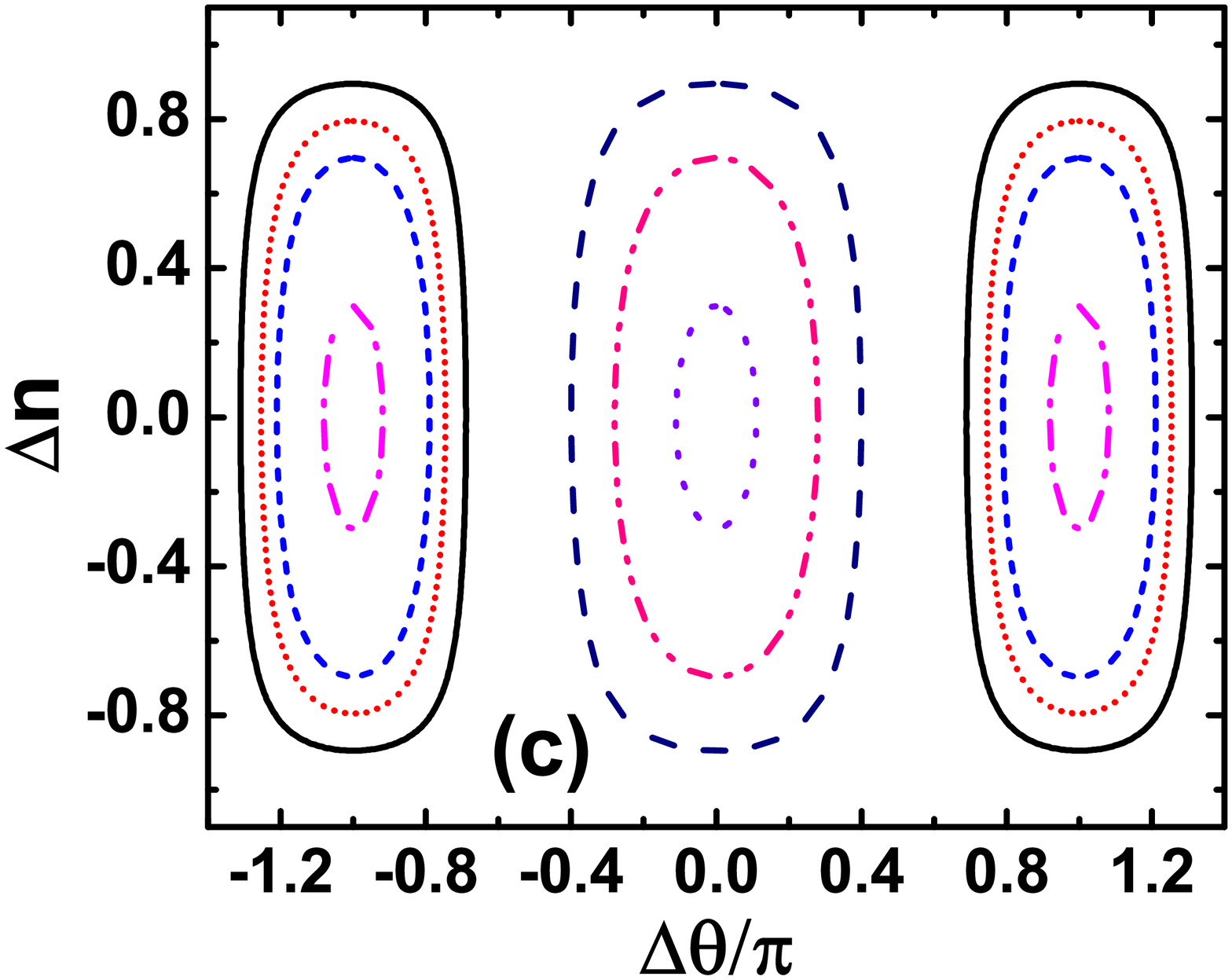}
\caption{(Color online) The $\Delta n$-$\Delta\theta$ phase space
diagram for different interaction strengths. (a) $g_{1D}=0.01$, (b)
$g_{1D}=0.005$, (c) $g_{1D}=0.001$. The orbits are presented by
constant energy lines. } \label{fig:ph-z}
\end{figure*}

In following calculations, we set $\beta=10$, $v_b=5$ and the
dimensionless trap geometry are chosen as
$v_{1D}(z)=z^2+10\mathrm{cos}^2(\pi z/5.2)$, where $q_0^{\prime}=5.2$
is calculated from the experiment parameter in Ref.
\protect\cite{Oberthaler2005}, $q_0\sim5\mu m$. However, the
dimensionless interacting strength $g_{1D}$ is a tunable parameter.
It varies with particle number of the system, $N$. The time step is
$\delta\tau=0.001$ and space step $\delta z=0.01$ in the
calculation.

The evolution of the population imbalance $\Delta n$ and the phase
difference $\Delta\theta$ can be calculated from Eq. (\ref{GP1D}).
Figure~\ref{fig:MQST} shows the obtained results for $g_{1D}=0.01$,
with the initial states being ($\Delta n=0.9$ $\Delta\theta=0$),
(0.9, $\pi$) and (0.5, $\pi$), respectively. The solid lines in
Fig.~\ref{fig:MQST}(a) and \ref{fig:MQST}(b) depict evolutions from
the state ($\Delta n=0.9$ $\Delta\theta=0$). Both the population
imbalance and the phase difference oscillate around zero ($\Delta
n=0$ and $\Delta\theta=0$) with time. This denotes the Josephson
oscillation mode. In the other two cases, $\Delta n$ is always
positive which means that the particle number in the left well is
always larger that in the right well. This is referred to as the
so-called macroscopic quantum self-trapping phenomenon
\cite{Smerzi1997}. As shown in Fig.~\ref{fig:MQST}(b), the short
dashed-line oscillates around $\Delta\theta=\pi$ while the short-dotted
line indicates that the phase changes monotonously with time. Such
tendency characterize the difference between the $\pi$-mode MQST
and running phase MQST.

By combining evolutions of $\Delta n$ and $\Delta\theta$, the
$\Delta n$-$\Delta\theta$ phase space diagram can be constructed.
Fig.~\ref{fig:ph-z}(a) displays the three kinds of evolutions
discussed above. (i) The Josephson oscillation mode characterized by
the time-average particle imbalance $\langle\Delta
n\rangle_{\tau}=0$ is represented by closed orbits around
$\Delta\theta=0$ (or $\Delta\theta=2k\pi$ where $k$ is an integer).
(ii) The $\pi$-mode MQST is represented by the closed orbits around
$\Delta\theta=\pi$ (or $\Delta\theta=(2k\pm1)\pi$). (iii) The open
orbits between the above two types of closed orbits refer to the
running-phase MQST. The particle imbalance $\langle\Delta
n\rangle_{\tau}\neq 0$ for both the two MQST modes.

The $\Delta n$-$\Delta\theta$ phase space diagram has been produced
previously based on the two-mode model \cite{Smerzi1997}. As Ref.
\protect\cite{Smerzi1997} suggested, each orbit denotes a constant
energy contour line. The $\pi$-mode has relative high energy. The
energy maxima lie at the center of the $\pi$-mode closed orbits and
there are two maxima symmetrically located on both sides of $\Delta
n=0$ axis. On the contrary, the Josephson oscillation mode has
smaller energy and the energy minimum is located at the points
($\Delta n=0$, $\Delta\theta=2k\pi$).

Nevertheless, it is suggested that the two-mode model has to be
improved to agree with experimental observations
\cite{Oberthaler2005}. Meanwhile, it is shown that the TDGP equation
provides more accurate simulation of dynamical evolution of the
double-well condensate than the two-mode model \cite{Bergeman2006, Sanpera2011}.
Unlike the two-mode model which ignores the spatial
information, the TDGP equation deals with the spatial distribution
of density and phase, thus it achieves quantitative accuracy of
dynamics of the condensate. The present study produces the $\Delta
n$-$\Delta\theta$ phase space diagram systematically within the
framework of TDGP equation.

MQST is a kind of nonlinear effect which arises because of the
interaction between bosons and might disappear if the interaction
becomes weaker. In Fig.~\ref{fig:ph-z}(b) where $g_{1D}=0.005$, the
running-phase mode disappears and the $\pi$-mode MQST is surrounded
by large closed circles referred as the $\pi$-mode Josephson
oscillation. In Fig.~\ref{fig:ph-z}(c) where $g_{1D}=0.001$, the
$\pi$-mode MQST disappears also.

\begin{figure}[htb]
\includegraphics[width=0.8\linewidth]{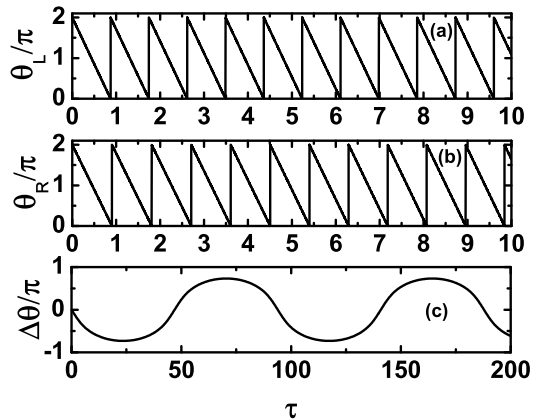}
\caption{Average phases in the left well, $\theta_L$ (a), the right
well, $\theta_R$ (b) as well as the phase difference $\Delta\theta$
(c) for a Josephson oscillation case corresponding to the
long-dashed line in Fig.~\ref{fig:MQST}a.} \label{fig:ph-l-r}
\end{figure}

$\Delta\theta$ is the conjugate variable of $\Delta n$. We can
calculate the average phase in each well according to Eqs.
(\ref{theta_L}) and (\ref{theta_R}). Fig.~\ref{fig:ph-l-r} shows
$\theta_L$, $\theta_R$ and $\Delta\theta$ for a Josephson
oscillation case. It is worth noting that the phase of the
condensate in each well varies much faster than the phase difference
between the two wells. The phase in each well changes almost
synchronously with slightly different high frequencies. In the
two-mode model, $\theta_L$ and $\theta_R$ are not defined.

\section{The dissipation effect}\label{sect:dissipation}

\begin{figure*}[htb]
\includegraphics[width=0.75\linewidth]{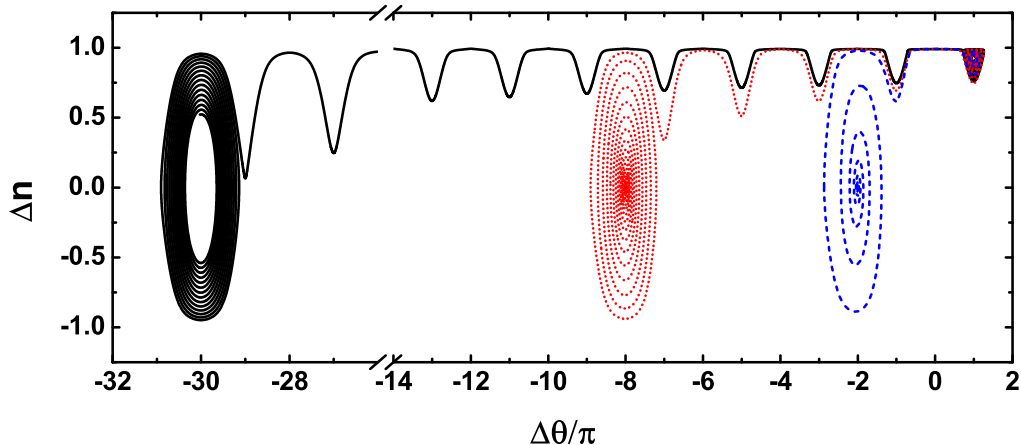}
\caption{(Color online) Evolution trajectories of a system initially
in the $\pi$-mode MQST state ($\Delta n=0.9$, $\Delta\theta=\pi$)
with different dissipation parameters, $\gamma=0.01$ (solid line),
$\gamma=0.03$ (short-dotted line), and $\gamma=0.1$ (short-dashed
line), respectively. } \label{fig:diss1}
\end{figure*}

To proceed, we investigate dynamical behaviors of the double-well
condensate subject to unavoidable dissipation. We consider a system
described by phase space diagram in Fig.~\ref{fig:ph-z}(a) and
suppose that it is initially in a $\pi$-mode MQST state,
($\Delta n=0.9$, $\Delta\theta=\pi$).

Fig.~\ref{fig:diss1} shows the evolution trajectories calculated
using Eq.~(\ref{GP1D-Dis}). The dissipation parameter is chosen to
be $\gamma=0.01$, $0.03$ and $0.1$, respectively. In all the three
cases, the trajectory passes through three different regimes. In the
$\pi$-mode MQST regime, the system moves along an outward spiral
path. Then it enters the running-phase MQST regime where the phase
difference changes significantly. At last, it slides into the
Josephson oscillation regime, along an inner spiral path. As
discussed above, the energy decreases from the $\pi$-mode MQST to
the Josephson oscillation. Therefore such trajectories imply that
energy dissipates gradually. Moreover, the larger the dissipation
parameter is, the faster the system drops into the lowest energy
state.

Marino {\it et al.} have extended the two-mode model to include the
dissipation effect \cite{Smerzi}. The dissipation term they employed
is based on the address in Ref.~\protect\cite{Leggett1998}, where it
was assumed to be an Ohmic current of normal atoms which causes
energy dissipation. Present study is qualitatively consistent with
results of Ref.~\protect\cite{Smerzi}. Since our approach is an
extension of the TDGP equation approach, we believe that the
obtained results provide a more quantitatively accurate description
of the dissipation process for the double-well condensate and are
expected to be compared with experiments, as discussed in
Sect.~\ref{sect:diagram}.

\begin{figure}[htb]
\includegraphics[width=0.75\linewidth]{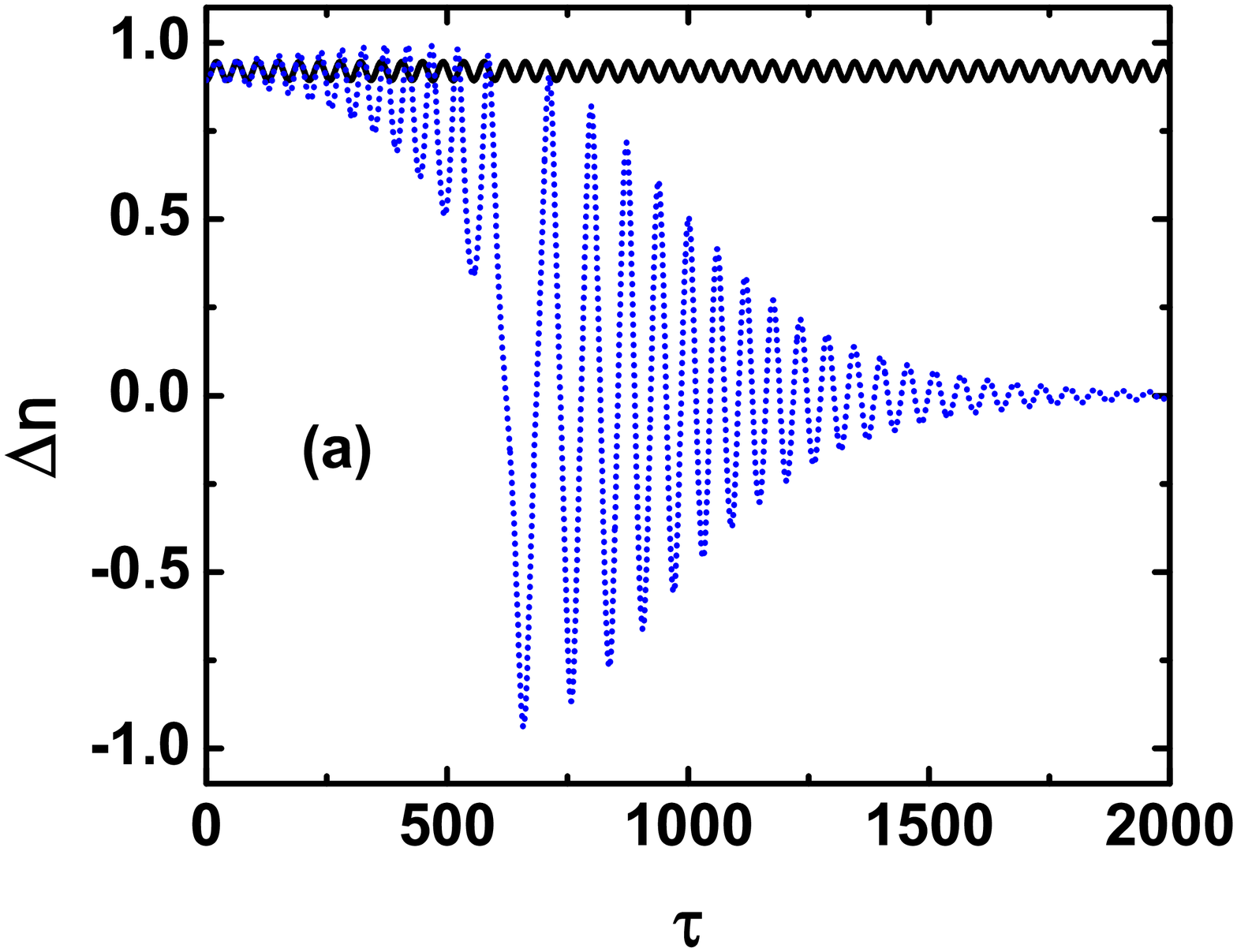}
\includegraphics[width=0.75\linewidth]{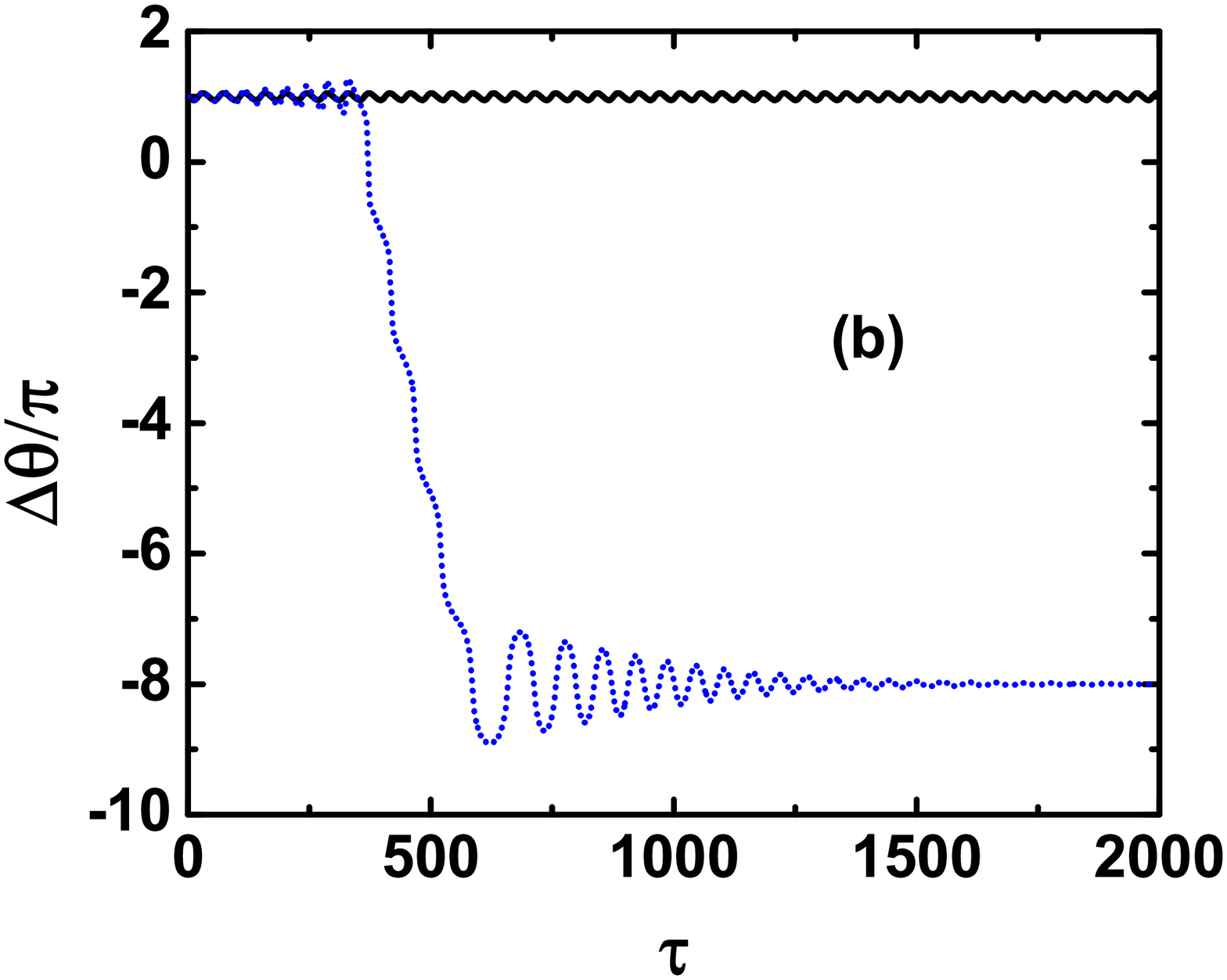}
\caption{(Color online) Evolution of $\Delta n$ (a) and
$\Delta\theta$ (b) starting from a $\pi$-mode MQST point ($\Delta
n=0.9$, $\Delta\theta=\pi$) with dissipation parameters
$\gamma=0.03$ (short-dotted lines), compared with case without
dissipation (solid lines).} \label{fig:diss2}
\end{figure}

To illustrate the dissipation process clearly, we calculate
time-evolutions of the population imbalance and the phase
difference, taking the $\gamma=0.03$ case as an example, as shown in
Fig.~\ref{fig:diss2}. When $\tau$ is less than $350$, the system is
still in the $\pi$-mode MQST regime where both $\Delta n$ and
$\Delta \theta$ oscillate around the center of the $\pi$-mode MQST
region, although their amplitudes grow slightly. During the interval
from $\tau\sim 350$ to $\tau\sim 600$, $\Delta \theta$ drops
abruptly from $\pi$ to $-8\pi$, which is the typical character of
the running-phase MQST regime. When $\Delta\theta$ starts to
oscillate around $-8\pi$, the system enters into the Josephson
oscillation regime. During this stage, both $\Delta n$ and $\Delta
\theta$ oscillate in a sine-like way, but their amplitudes damp with
time. Finally, $\Delta n$ arrives at the value of zero which
represents an equilibrium distribution of particles in the two
wells.

\begin{figure}[htb]
\includegraphics[width=0.95\linewidth]{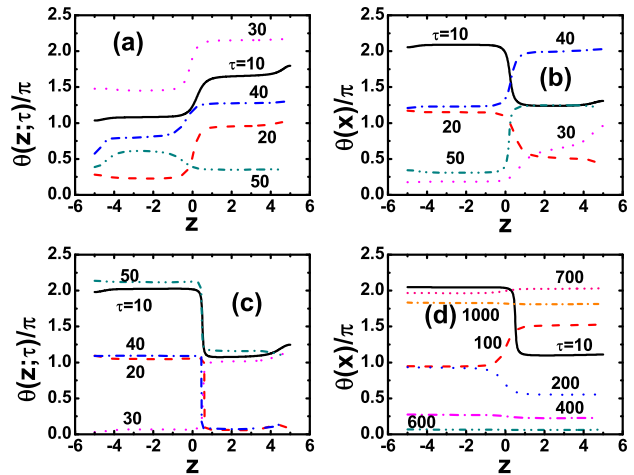}
\caption{(Color online) Snapshots of phase distributions in the
double-well at different given times. (a), (b) and (c) depict the
Josephson oscillation, running-phase MQST and $\pi$-mode MQST cases
without dissipation, respectively. (d) shows the dissipation prosess
from the $\pi$-mode MQST state with $\gamma=0.03$.}
\label{fig:ph-distr}
\end{figure}

\begin{figure}[htb]
\includegraphics[width=0.75\linewidth]{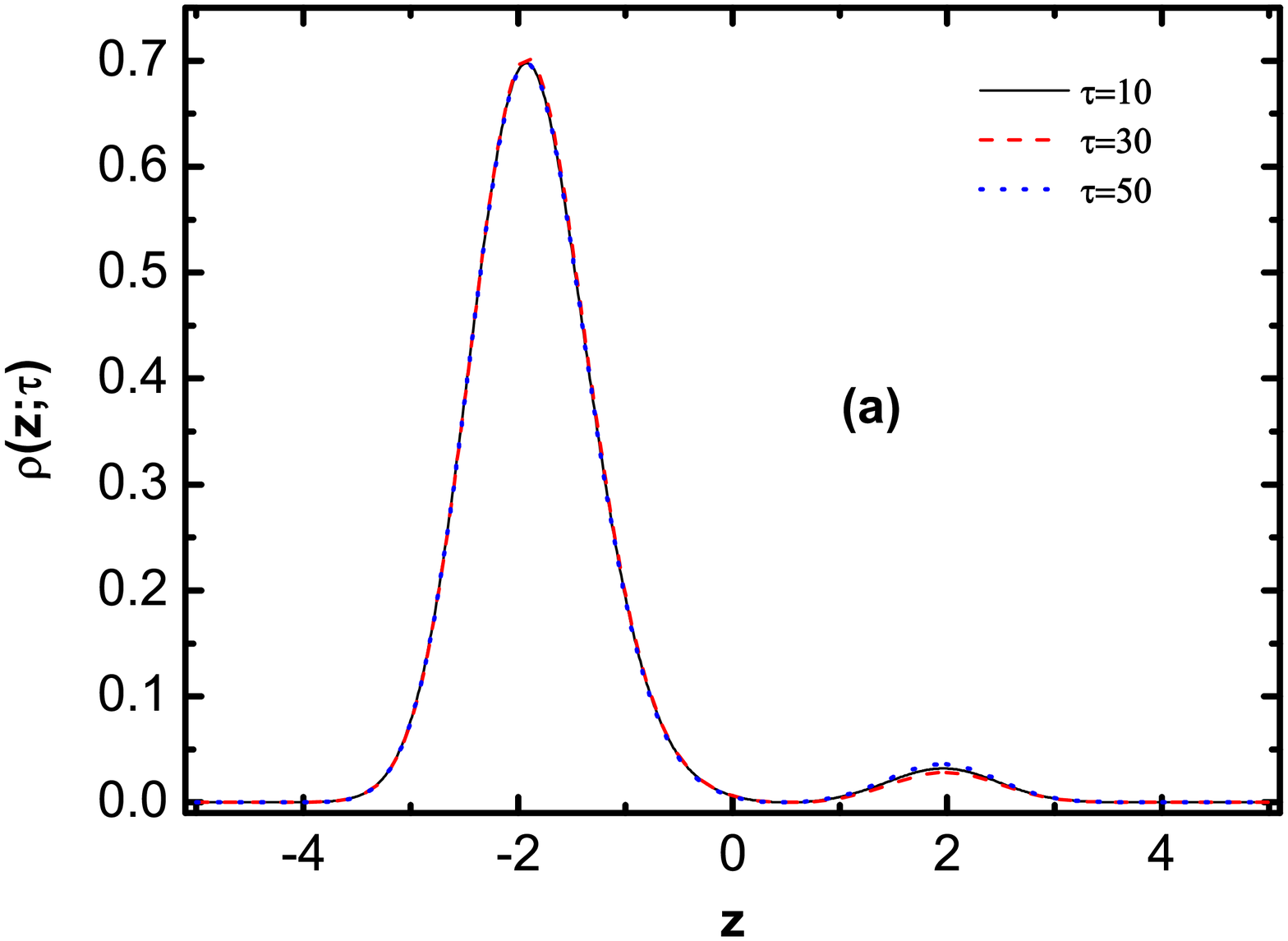}
\includegraphics[width=0.75\linewidth]{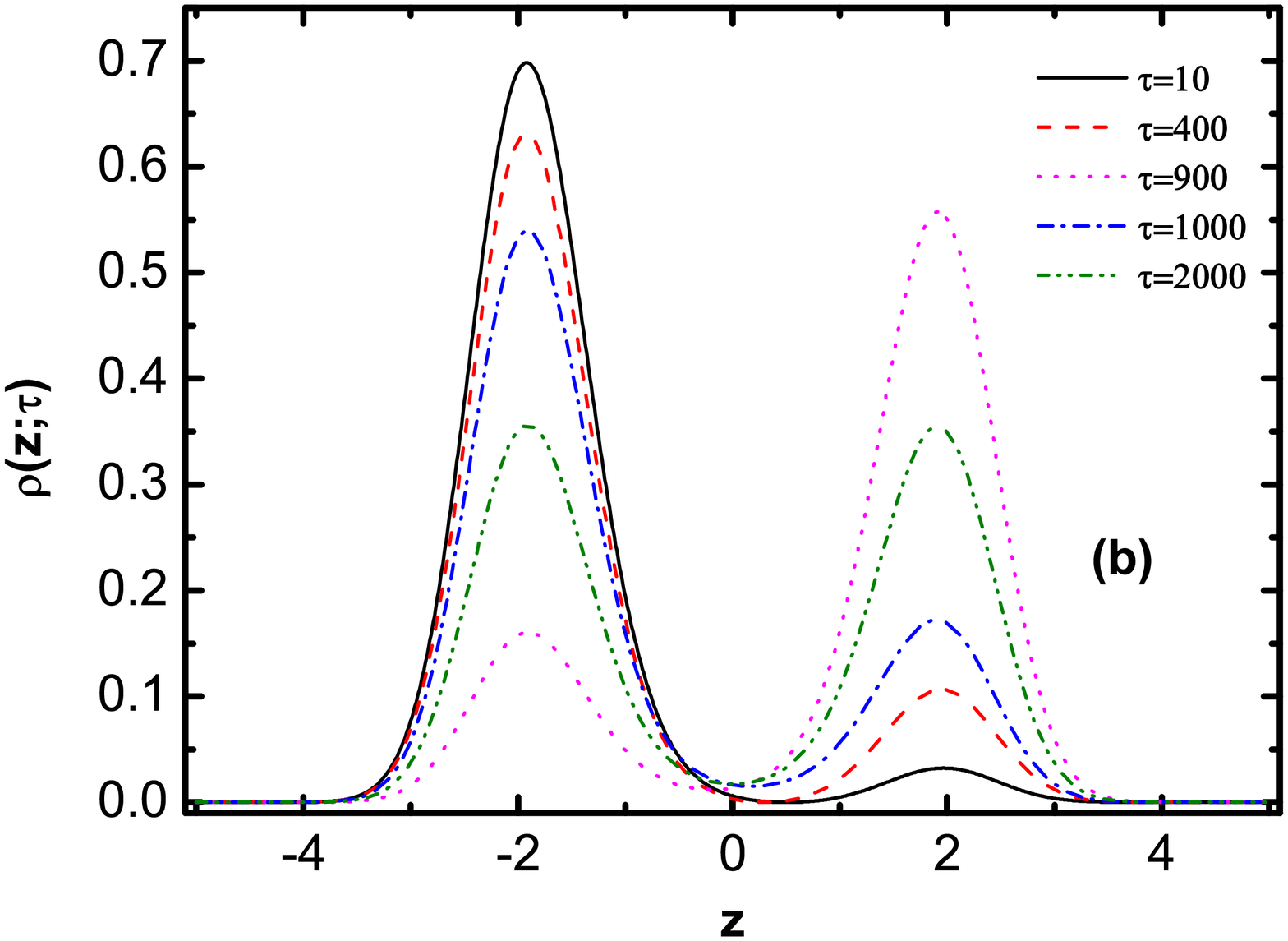}
\caption{(Color online) Snapshots of particle density distributions
in the double-well. (a) depicts the $\pi$-mode MQST case without
dissipation. (b) shows the dissipation prosess from the $\pi$-mode
MQST state with $\gamma=0.03$.} \label{fig:diss3}
\end{figure}

An advantage of the TDGP equation approach is that it provides
spatial information of the phase distribution in each well
\cite{Bergeman2006}. Fig.~\ref{fig:ph-distr} illustrates the phase
distribution, $\theta(z;\tau)$, for different dynamical processes.
In all the cases, the phase keeps almost a constant in each well,
except at the edges of the well. Even if the phase varies with the
position, the deviation is usually very small. This accounts for the
validity of the use of two-mode approximation. An interesting
observation is that the phase distribution in the well with more
particles seems flatter than that in the well with less particles.
This point is demonstrated in Fig.~\ref{fig:ph-distr}(b) and
\ref{fig:ph-distr}(c) which depict the MQST cases with the left well
trapping more particles. We also find that the phase distribution in
the dissipative case is much flatter, as if the dissipation tends to
suppress phase fluctuation in each well, as shown in
Fig.~\ref{fig:ph-distr}(d).

Fig.~\ref{fig:ph-distr} also shows that the phase $\theta(z;\tau)$
may change abruptly near the barrier, but not right at $z=0$. Taking
Fig.~\ref{fig:ph-distr}(c) which describes the $\pi$-mode MQST as an
example, there is an insanely steep gradient near $z\approx 0.5$ on
each lines.

We now look at the particle density distribution.
Fig.~\ref{fig:diss3}(a) displays the $\pi$-mode MQST case without
dissipation, where $n(z)$ exhibits two peaks located in the two
wells respectively. Obviously, the left peak keeps higher than the
right one all the time. We note that the minimum between the two
peaks does not lie at the center of the barrier. It is biased
towards the right side of the barrier. This accounts for the reason
why $\theta(z;\tau)$ begins to change abruptly on right side of the
barrier. We judge that the abrupt change in $\theta(z;\tau)$ takes
place at the minimum point of the particle density distribution
between the two density peaks. Fig.~\ref{fig:diss3}(b) shows the
dissipation process from the $\pi$-mode MQST with $\gamma=0.03$.
Then the system evolves into the running-phase MQST regime (see the
$\tau=400$ line) and the Josephson oscillation regime (see
$\tau=900$, $1000$ and $2000$ lines). At $\tau=2000$, the system
becomes almost stable.


\section{Conclusion}\label{sect:conclusion}

In conclusion, we have investigated dynamic properties of the
double-well Bose-Einstein condensate subject to dissipation in the
framework of time-dependent Gross-Pitaevskii equation. The phase
space diagram for the system without dissipation are systematically
produced and three typical dynamical regimes of Josephson
oscillation, $\pi$-mode MQST and running phase MQST are shown
clearly. The energy dissipation processes are studied by introducing
a phenomenological dissipation term to the TDGP equation. The
evolution trajectories evolve gradually from the high energy states
to the low energy states, suggesting that this approach is
applicable to describing energy dissipation effect. Relevant results
have been studied previously based on two-mode model. We expect that
the TDGP approach provides a more quantitatively accurate
description of the dissipation process for the double-well
condensate and the results are expected to be compared with
experiments.

Moreover, the TDGP approach produces spatial information of the
phase distribution and density profile which are not directly
reflected in two-mode model calculations. The phase of condensate
varies slowly in space in each well and may exhibit abrupt changes
near the barrier, but not certainly at the barrier center. These
sudden changes occur at the minimum position in particle density
profile. We also note that the average phase in each well varies
much faster with time than the phase difference between two wells.

\begin{acknowledgments}
This work is supported by the National Natural Science Foundation
of China (Grant No. 11074021 and Grant No. 11004007), and the
Fundamental Research Funds for the Central Universities of China.
\end{acknowledgments}

\end{document}